\documentclass[aps,prx,twocolumn,unsortedaddress,floatfix,nofootinbib,superscriptaddress]{revtex4-2}
\usepackage{amsthm}
\usepackage{amsfonts}
\usepackage{siunitx}
\usepackage{amsmath}
\usepackage{tabularx}
\usepackage{amssymb}
\usepackage{graphicx}
\usepackage{verbatim}
\usepackage[colorlinks]{hyperref}
\usepackage{tikz}
\usepackage{braket}
\usepackage{xcolor}
\usepackage{multirow}
\usepackage{booktabs}
\makeatletter
\def\@ssect@ltx#1#2#3#4#5#6[#7]#8{%
  \def\H@svsec{\phantomsection}%
  \@tempskipa #5\relax
  \@ifdim{\@tempskipa>\z@}{%
    \begingroup
    \interlinepenalty \@M
    #6{%
      \@ifundefined{@hangfroms@#1}{\@hang@froms}{\csname @hangfroms@#1\endcsname}%
      {\hskip#3\relax\H@svsec}{#8}%
    }%
    \@@par
    \endgroup
    \@ifundefined{#1smark}{\@gobble}{\csname #1smark\endcsname}{#7}%
  }{%
    \def\@svsechd{%
      #6{%
        \@ifundefined{@runin@tos@#1}{\@runin@tos}{\csname @runin@tos@#1\endcsname}%
        {\hskip#3\relax\H@svsec}{#8}%
      }%
      \@ifundefined{#1smark}{\@gobble}{\csname #1smark\endcsname}{#7}%
      \addcontentsline{toc}{#1}{\protect\numberline{}#8}%
    }%
  }%
  \@xsect{#5}%
}%
\makeatother

\definecolor{linkcolor}{RGB}{0,83,166}
\hypersetup{
  colorlinks = true,
  allcolors = {linkcolor}
}

\begin{document}
\newcommand{\mytitle}{A comment on comparing optimization on D-Wave and IBM quantum processors}
\newcommand{\affildw}{D-Wave Quantum, Burnaby, British Columbia, Canada}

\author{Catherine C.~McGeoch}
\affiliation{\affildw}
\author{Kevin Chern}
\affiliation{\affildw}
\author{Pau Farr\'{e}}
\affiliation{\affildw}
\author{Andrew D.~King}
\email[]{aking@dwavesys.com}
\affiliation{\affildw}

\title{\mytitle}

\date{\today}
\begin{abstract}
  Recent work \cite{Sachdeva2024} presented an iterative hybrid quantum variational optimization algorithm designed by Q-CTRL and executed on IBM gate-based quantum processing units (QPUs), claiming a significant performance advantage against a D-Wave{\texttrademark}  quantum annealer.  Here we point out major methodological problems with this comparison.  Using a simple unoptimized workflow for quantum annealing, we show success probabilities multiple orders of magnitude higher than those reported by Ref.~\cite{Sachdeva2024}.  These results, which can be reproduced using open-source code and free trial access to a D-Wave quantum annealer, contradict Q-CTRL's claims of superior performance.  We also provide a direct comparison between quantum annealing and a recent demonstration~\cite{Miessen2024} of digitized quantum annealing on an IBM processor, showing that analog quantum annealing on a D-Wave QPU reaches far lower energies than digitized quantum annealing on an IBM QPU.
\end{abstract}

\maketitle

\def\title#1{\gdef\@title{#1}\gdef\THETITLE{#1}}

\begin{table*}
  \begin{tabular*}{.8\textwidth}{@{\extracolsep{\fill}}cc|lcr|lcr}
    \toprule
    \multicolumn{5}{l}{\textbf{Max-cut}} \\
    \midrule
    \multicolumn{2}{l}{\textbf{Global}} & \multicolumn{3}{l}{\textbf{Q-CTRL (IBM Quantum)}} & \multicolumn{3}{l}{\textbf{QA (D-Wave)}} \\
    \multicolumn{2}{l}{} & \multicolumn{3}{l}{} & \multicolumn{3}{l}{Embedding time included} \\
    \midrule
    Instance & Opt & $P_\text{GS}$ & $t_\text{sample}$ (ms) & TTS (ms) & $P_\text{GS}$ & $t_\text{sample}$ (ms) & TTS (ms) \\
    \midrule
    (28,3,102,u)   & 40 & 0.94 & 25  & 40 & 1 & 1.2 & 1.2 \\
    (30,3,264,u)   & 43 & 0.92 & 25 & 46  & 1 & 1.2 & 1.2 \\
    (32,3,7,u)     & 46 & 0.83 & 25  & 66 & 1 & 1.3 & 1.3 \\
    (80,3,68,u)    & 106& 0.14 & 28 & 868 & 0.23 & 1.6 & 28 \\
    (100,3,12,u)   & 135& 0.12 & 28 & 1000 & 0.53 & 1.6 & 9.5 \\
    (120,3,8,u)    & 163& 0.09 & 28 & 1436 & 0.79 & 1.6 &  4.6 \\
    \midrule
    \midrule
    \multicolumn{5}{l}{\textbf{Higher-order spin glass}} \\
    \midrule
    \multicolumn{2}{l}{\textbf{Global}} & \multicolumn{3}{l}{\textbf{Q-CTRL (IBM Quantum)}} & \multicolumn{3}{l}{\textbf{QA (D-Wave)}}\\
    \midrule
    Instance & Opt & $P_\text{GS}$ & $t_\text{sample}$ (ms) & TTS (ms) &  $P_\text{GS}$ & $t_\text{sample}$ (ms) & TTS (ms)\\
    \midrule
    0  & -200     & 0.000134  & 28 & 962,210  & 0.002 &0.21& 488 \\
    3  & -198     & 0.12  &28 & 1027           & 0.018 &0.21& 53 \\
    5  & -198     & 0.20  & 28 & 568           & 0.048 &0.21& 20  \\
    10 & -202     & 0.019  &28 &  6651          & 0.135  &0.21& 6.7 \\
    11 & -180     & 0 & 28 & $\infty$           & 0.006 &0.21& 157 \\
    69 & -190     & 0 & 28 & $\infty$           & 0.009 &0.21& 110 \\
    \bottomrule
    
  \end{tabular*}
  \caption{
    \label{table:1}
    Comparison between \mbox{Q-CTRL} and QA on (top) unweighted max-cut and (bottom) higher-order spin-glass problems.  Compare with Ref.~\cite{Sachdeva2024} Table I.  All D-Wave experiments are run on Advantage\_System4.1 using 500 reads per programming and $\SI{1}{ms}$ annealing time.   Higher-order spin-glass inputs are constructed as in Ref.~\cite{Pelofske2024}, and use six parallel embeddings to draw six samples per read.  To reduce uncertainty, QA experiments were repeated five times for each higher-order spin-glass input, yielding 15,000 samples, comparable to \mbox{Q-CTRL} experiments.}
\end{table*}

Among the proposed approaches for accelerating optimization with quantum computers, quantum annealing (QA) and the quantum approximate optimization algorithm (QAOA) have attracted the most interest.  These approaches are run on annealing-based and gate-based quantum computers, respectively.  QAOA has previously been compared against QA and random sampling \cite{Willsch2020,Pelofske2021,Ushijima-Mwesigwa2021,Harrigan2021,Pelofske2024}.

One recent work from \mbox{Q-CTRL} \cite{Sachdeva2024} presents an iterative algorithm that is based on QAOA but which adds per-qubit angles $\theta_1,\ldots,\theta_N$ that bias the initial qubit states during the iterative optimization.  These angles $\theta_i$ play the role of linear-bias memory registers that are variationally tuned by the ansatz.  Setting aside how these biases affect the computational role of quantum effects in this algorithm, we simply remark that the hybrid approach departs significantly from standard QAOA; to avoid confusion, we refer to the algorithm as ``\mbox{Q-CTRL}''.

The matter we address here is the comparison, by Sachdeva {\it et al.}~\cite{Sachdeva2024}, between \mbox{Q-CTRL} and QA as implemented on a D-Wave quantum annealer.  Ref.~\cite{Sachdeva2024} claims significantly superior performance of \mbox{Q-CTRL} over QA, based on data taken from Pelofske {\it et al.}~\cite{Pelofske2024}.  However, there are several major methodological problems with this comparison, which we now briefly list.

First, QA success probabilities are calculated using the sum total of all runs performed in an exploratory parametric grid search.  The vast majority of the runs are obviously substandard {\em a priori}, and including them as serious problem-solving efforts results in much poorer statistical outcomes than would be produced by a basic, good-faith choice of parameters.  No analogous parametric sweep was included in \mbox{Q-CTRL}'s reported results.

Second, success probabilities are used as the sole figure of merit, without comparing the time spent producing solutions.  It is nonsensical to compare two algorithms with potentially vastly different runtimes using only success probability.  For example, this would make it impossible to outperform brute-force algorithms.  Time to solution (TTS) is a widely used metric~\cite{Roennow2014,Albash2016a,Cain2023,Bauza2024} that incorporates both runtime and success probability, and we use it here.

Third, classical postprocessing is applied, but only to \mbox{Q-CTRL} results.  This obscures the computational role of the variational ansatz itself in the \mbox{Q-CTRL} results, and does not afford QA the same benefit of bitflip error correction.

Fourth, instead of running on an arguably unbiased set of six inputs, Ref.~\cite{Sachdeva2024} compared \mbox{Q-CTRL} vs.~QA using the three worst QA results from a 10-input testbed from Ref.~\cite{Pelofske2024}, and three other auxiliary inputs also from Ref.~\cite{Pelofske2024} for which no QA results were published.

\section{Input types}

Here we consider two types of optimization problems expressed in the classical Ising model of $\pm 1$ spins.

Ref.~\cite{Sachdeva2024} compares Q-CTRL and D-Wave on higher-order unconstrained binary optimization problems from Ref.~\cite{Pelofske2024}, whose linear, quadratic, and cubic Ising energy terms are laid out in a 127-qubit ``heavy-hex'' geometry used by IBM gate-model processors.  These inputs are specifically tailored to be amenable to both QAOA on IBM processors and QA on D-Wave Advantage{\texttrademark} processors.  The cubic terms somewhat favor QAOA because in QA they must be order-reduced via the introduction of 138 slack variables in addition to the 127 spins in the system.  We construct the instances and run the QA experiments as in Ref.~\cite{Pelofske2024}, highlighting the fact that \mbox{Q-CTRL}'s claimed advantage disappears when standard and even-handed benchmarking methods are applied (Table~\ref{table:1}).

Ref.~\cite{Sachdeva2024} also published results on weighted and unweighted max-cut problems.  The unweighted problems are generated by NetworkX~\cite{networkx} with an explicit random seed, so we can run QA on these as well.  Table~\ref{table:1} presents results from five such problems, named $(N,d,s,u)$ where $N$, $d$, and $s$ are the graph size, degree, and random seed, respectively, and ``u'' indicates an unweighted max-cut problem.  Again, we will see that QA outperforms \mbox{Q-CTRL}.

Additionally, we note that a better choice of gadgets than the one used in Ref.~\cite{Pelofske2024}\footnote{This better choice simply replaces the upper gadget for energy penalty $d_{BAC}=-1$  in Ref.~\cite{Pelofske2024} Fig.~3 with the upper gadget for $d_{BA(-C)}=+1$.} leads to significantly improved QA results (see Table~\ref{table:2}).

\section{Main results}

Instead of attempting to optimize QA parameters, we run QA on all inputs using 500 reads per QPU call, annealing time $t_a=\SI{1}{ms}$, and all other parameters at default values.  We run the same postprocessing on QA results as on Q-CTRL results, and we examine the effect of this postprocessing in the next section.

Table~\ref{table:1} summarizes our experimental results.  To estimate \mbox{Q-CTRL} time per sample $t_\text{sample}$, we refer to Ref.~\cite{Sachdeva2024} Appendix A, which specifies 6000 samples taking 150 seconds of QPU time for the three smallest inputs, and larger inputs taking 7 minutes for 15,000 samples; thus, 25 and 28 ms for these problem classes respectively.

Max-cut problems on random graphs cannot be assumed to fit directly into the Pegasus qubit connectivity graph of the D-Wave Advantage processor, so we heuristically minor-embed them using D-Wave's Ocean{\texttrademark} software development kit~\cite{Ocean}, and include this overhead in the total runtime.

The higher-order spin-glass problems are specifically tailored to fit into both IBM and D-Wave hardware, and we use the same six parallel embeddings for all inputs.  Thus QA yields 3000 samples for each 500-read QA call, and $t_\text{sample}$ is less than the annealing time of $\SI{1}{ms}$ for these problems, even when QPU overhead is included.

Given that QA produces samples far faster than \mbox{Q-CTRL}, we compare the approaches on equal footing using the time-to-solution (TTS) metric, which estimates the time required to find a ground state with 99\% probability.  For a per-sample time $t_\text{sample}$ and a ground-state probability $P_\text{GS}$,
\begin{equation}
  \text{TTS} = t_\text{sample}\cdot\max\left( \frac{\log{(1-0.99)}}{\log{(1-P_\text{GS})}}, 1\right).
\end{equation}
The max-cut inputs are easy enough to be solved with greedy descent: Ref.~\cite{Sachdeva2024} reports that their ``local solver'' finds ground states for all such inputs; this local solver finds five local minima from a random initial state using greedy bitflips, and takes the best solution of the five as its output.

On all but two instances (Table~\ref{table:1}), QA shows better success probability than \mbox{Q-CTRL}.  QA shows better time to solution (TTS) on all inputs, by a minimum of $30\times$.

\begin{figure}
\includegraphics{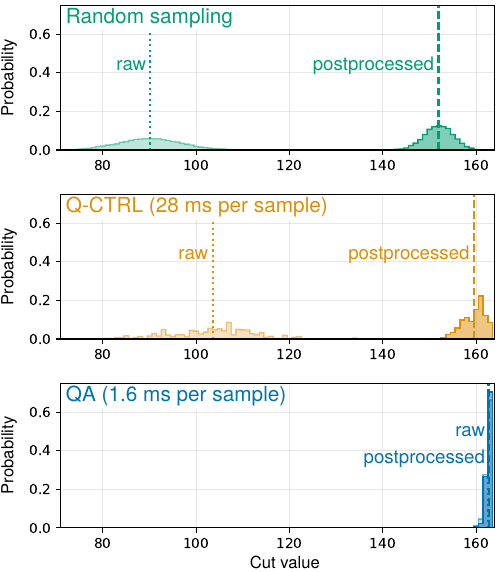}
  \caption{Effect of postprocessing on max-cut instance (120,3,8,u).  We show raw and postprocessed solution quality histograms for random sampling (top), \mbox{Q-CTRL} (middle) and QA (bottom).  Dotted and dashed lines indicate raw and postprocessed mean values, respectively.  \mbox{Q-CTRL} data and postprocessing method are taken from Ref.~\cite{Sachdeva2024}.  Optimal value of $163$ was achieved by postprocessed \mbox{Q-CTRL} and both raw and postprocessed QA.}\label{fig:pp}
\end{figure}

\section{Effect of postprocessing}

Ref.~\cite{Sachdeva2024} shows \mbox{Q-CTRL} results without classical postprocessing for only one input: max-cut (120,3,8,u).  We can therefore compare the effect of postprocessing on \mbox{Q-CTRL} and QA on this input.  The postprocessing applied here sweeps through the spins in random order a maximum of five times, and flips a spin when it strictly improves the solution, thus finding a local minimum.  Fig.~\ref{fig:pp} shows data for raw and postprocessed samples from random sampling, \mbox{Q-CTRL}, and QA.  The data clearly shows that classical postprocessing is crucial to achieving good samples with \mbox{Q-CTRL}, and is far less important in QA.

Moving to the higher-order spin-glass problems, Table~\ref{table:2} compares raw and postprocessed ground-state probabilities for QA.  We show the effect of postprocessing for both the original input construction of Pelofske {\it et al.}~\cite{Pelofske2024}---as considered in Ref.~\cite{Sachdeva2024} and our Table~\ref{table:1}---and a modified construction in which the third gadget from \cite{Pelofske2024} Fig.~3 is replaced by a version of the first gadget, with an appropriate spin-reversal transformation applied.  This modification doubles the QA energy scale and significantly reduces the need for postprocessing.  Forthcoming Advantage2{\texttrademark} systems  have roughly a 2x relative boost to energy scales compared to the Advantage processors used here.  An Advantage2 prototype is available online and can be used with free trial access to the Leap{\texttrademark} quantum cloud service~\cite{Leap}.

\begin{table*}[!t]
  \begin{tabular*}{.9\linewidth}{@{\extracolsep{\fill}}c|ll|rr|ll|rr} %
    \toprule
    \multicolumn{9}{l}{\textbf{Higher-order spin glass}} \\
    \midrule
    & \multicolumn{4}{l|}{\textbf{QA}}  & \multicolumn{4}{l}{\textbf{QA}} \\
    & \multicolumn{4}{l|}{Construction from \cite{Pelofske2024}}  & \multicolumn{4}{l}{Better gadgets} \\
    \midrule
     &   \multicolumn{2}{c|}{$P_\text{GS}$}  &  \multicolumn{2}{c|}{TTS (ms)}&\multicolumn{2}{c|}{$P_\text{GS}$}   & \multicolumn{2}{c}{TTS (ms)}\\
\midrule
    Inst. & raw\hspace{8mm} & postprocessed & \hspace{8mm}raw & postprocessed & raw\hspace{8mm} & postprocessed & \hspace{8mm}raw & postprocessed \\
    \midrule
    0  & 0.0005 & 0.002  &1813& 488 & 0.067 & 0.073  &14& 13 \\
    3  & 0.005  & 0.018  &193& 53 & 0.22  & 0.24   &4.0& 3.6 \\
    5  & 0.003  & 0.048  &302& 20  & 0.42  & 0.57   &1.8& 1.1 \\
    10 & 0.087  & 0.135  &11& 6.7 & 0.58  & 0.61   &1.1& 1.0 \\
    11 & 0.002  & 0.006  &456& 157 & 0.12  & 0.15   &7.6& 6.1 \\
    69 & 0.002  & 0.009  &483& 110 & 0.07  & 0.08   &13& 11 \\
    \bottomrule
  \end{tabular*}
  \caption{\label{table:2}Effect of postprocssing on QA success probabilities and time to solution.  Right-hand data (``better gadgets'') replaces the third gadget from Ref.~\cite{Pelofske2024} Fig.~3 with a spin-reversal transform of the first gadget.  This doubles the QA energy scale, improving peformance significantly and making postprocessing less impactful.}
\end{table*}

\section{Comparing digitized and analog QA between qubit modalities}

\begin{figure}
\includegraphics{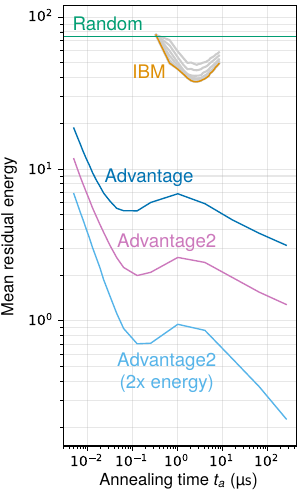}
\caption{Comparison of IBM and D-Wave processors on a 133-qubit heavy-hex spin-glass optimization problem.  Residual energy (distance from optimal state) is shown as a function of annealing time.  D-Wave Advantage system and Advantage2 prototype run analog quantum annealing; IBM Torino system runs digitized quantum annealing.  Annealing times for IBM are based on an assumption of 4-gate depth per Trotter slice, with $\SI{84}{ns}$ gate time.}\label{fig:sg}
\end{figure}

Recent work \cite{Miessen2024} evaluated the quality of gate-model quantum processors using digitized quantum annealing, within a framework of quantum critical dynamics.  Similar evaluations have been demonstrated on D-Wave (analog) QA processors \cite{Bando2020,King2022,King2023, Amin2023}.  In particular, Ref.~\cite{Miessen2024} demonstrated digitized QA on a 133-qubit heavy-hex spin glass.  This planar spin glass has couplings $J_{ij}$ distributed uniformly in $[-1,1]$.  This provides an opportunity to directly compare two quantum-computing modalities for optimization, with no hybrid methods, preprocessing, or postprocessing obscuring the contribution of the QPU.  We show such a comparison in Fig.~\ref{fig:sg}.  Residual energies are much lower for QA, particularly for the Advantage2 prototype, owing to higher energy scale and lower noise.

We also test the effect of doubling energy scale in the Advantage2 prototype.  Programmed couplings $J_{ij}$ are constrained to the interval $[-2,1]$, but this problem is small and sparse, and the couplings with $|J_{ij}|>0.5$ induce a subgraph with no cycles.  We can therefore find a spin-reversal transformation in linear time that compresses the spin-glass couplings to $[-1,1/2]$ by flipping the sign of a subset of spin variables.  This allows us to double the energy scale, leading to significantly smaller residual energies.  This is informative but is not a method that generalizes well to large, dense problems.

Finally we remark that this spin-glass input is not computationally challenging, but these results shed light on the nature of quantum annealing in gate-model and annealing-based QPUs.

\section{Conclusions}

We have scrutinized recent claims~\cite{Sachdeva2024} of gate-model optimization outperforming quantum annealing.  These claims were founded on a benchmarking methodology that mistook poor-quality samples from a broad parameter sweep~\cite{Pelofske2024} for a good-faith effort to optimize efficiently, and ignored the cost of producing a sample, instead measuring only sample quality.  Accounting for these issues, we found quantum annealing running on D-Wave processors to outperform \mbox{Q-CTRL}'s hybrid variational algorithm running on IBM quantum processors.  Our experiments can be repeated using free trial QPU access via D-Wave's Leap cloud service~\cite{Leap}.

We also compared solution quality on a planar spin-glass instance, and found that digitized quantum annealing running on IBM quantum processors is not competitive with analog quantum annealing running on D-Wave processors.

\section{Code availability}

Supporting data and code for running these experiments is available at Zenodo repository \url{https://doi.org/10.5281/zenodo.12549342}.

\section{Acknowledgments}

We are grateful to Elijah Pelofske, Alex Miessen and Guglielmo Mazzola for helpful discussions and providing experimental data.

\bibliography{paper}

\end{document}